\documentclass[twocolumn]{aastex63}
\usepackage{multirow}
\usepackage{amsmath}
\usepackage{booktabs}

\usepackage{xcolor}

\newcommand\lsim{\mathrel{\rlap{\lower4pt\hbox{\hskip1pt$\sim$}}
\raise1pt\hbox{$<$}}}
\accepted{ApJ}

\shorttitle{Star Formation with Streaming}
\shortauthors{Lake et al.}
\graphicspath{{./}{figures/}}
\begin{document}

\title{ The Supersonic Project: Early Star Formation with the Streaming Velocity}


\correspondingauthor{William Lake}
\email{wlake@astro.ucla.edu}
\author[0000-0002-4227-7919]{William Lake}
\affil{Department of Physics and Astronomy, UCLA, Los Angeles, CA 90095}
\affil{Mani L. Bhaumik Institute for Theoretical Physics, Department of Physics and Astronomy, UCLA, Los Angeles, CA 90095, USA\\}

\author[0000-0003-2369-2911]{Claire E. Williams}
\affil{Department of Physics and Astronomy, UCLA, Los Angeles, CA 90095}
\affil{Mani L. Bhaumik Institute for Theoretical Physics, Department of Physics and Astronomy, UCLA, Los Angeles, CA 90095, USA\\}

\author[0000-0002-9802-9279]{Smadar Naoz}
%\author{Smadar Naoz}
%\affil{Department of Physics and Astronomy, University of California, Los Angeles, CA 90095\\}
\affil{Department of Physics and Astronomy, UCLA, Los Angeles, CA 90095}
\affil{Mani L. Bhaumik Institute for Theoretical Physics, Department of Physics and Astronomy, UCLA, Los Angeles, CA 90095, USA\\}

\author[0000-0003-3816-7028]{Federico Marinacci}
%\author{Federico Marinacci}
\affiliation{Department of Physics \& Astronomy ``Augusto Righi", University of Bologna, via Gobetti 93/2, 40129 Bologna, Italy\\}
\affiliation{INAF, Astrophysics and Space Science Observatory Bologna, Via P. Gobetti 93/3, 40129 Bologna, Italy\\}

\author[0000-0001-5817-5944]{Blakesley Burkhart}
%\author{Blakesley Burkhart}
\affiliation{Department of Physics and Astronomy, Rutgers, The State University of New Jersey, 136 Frelinghuysen Rd, Piscataway, NJ 08854, USA \\}
\affiliation{Center for Computational Astrophysics, Flatiron Institute, 162 Fifth Avenue, New York, NY 10010, USA \\}

\author[0000-0001-8593-7692]{Mark Vogelsberger}
%\author{Mark Vogelsberger}
\affil{Department of Physics and Kavli Institute for Astrophysics and Space Research, Massachusetts Institute of Technology, Cambridge, MA 02139, USA\\}

\author[0000-0001-7925-238X]{Naoki Yoshida}
\affiliation{Department of Physics, The University of Tokyo, 7-3-1 Hongo, Bunkyo, Tokyo 113-0033, Japan}
\affiliation{Kavli Institute for the Physics and Mathematics of the Universe (WPI), UT Institute for Advanced Study, The University of Tokyo, Kashiwa, Chiba 277-8583, Japan}
\affiliation{Research Center for the Early Universe, School of Science, The University of Tokyo, 7-3-1 Hongo, Bunkyo, Tokyo 113-0033, Japan}

\author[0000-0001-6246-2866]{Gen Chiaki}
\affiliation{Astronomical Institute, Tohoku University, 6-3, Aramaki, Aoba-ku, Sendai, Miyagi 980-8578, Japan}

\author[0000-0002-8859-7790]{Avi Chen}
\affiliation{Department of Physics and Astronomy, Rutgers, The State University of New Jersey, 136 Frelinghuysen Rd, Piscataway, NJ 08854, USA \\}

\author[0000-0003-4962-5768]{Yeou S. Chiou}
\affil{Department of Physics and Astronomy, UCLA, Los Angeles, CA 90095}
\affil{Mani L. Bhaumik Institute for Theoretical Physics, Department of Physics and Astronomy, UCLA, Los Angeles, CA 90095, USA\\}



\begin{abstract}

At high redshifts ($z\gtrsim12$), the relative velocity between baryons and dark matter (the so-called streaming velocity) significantly affects star formation in low-mass objects. Streaming substantially reduces the abundance of low-mass gas objects while simultaneously allowing for the formation of supersonically-induced gas objects (SIGOs) and their associated star clusters outside of dark matter halos. Here, we present a study of the population-level effects of streaming on star formation within both halos and SIGOs in a set of simulations with and without streaming. Notably, we find that streaming actually enhances star formation within individual halos of all masses at redshifts between $z=12$ and $z=20$. This is demonstrated both as an increased star formation rate per object as well as an enhancement of the  Kennicutt-Schmidt relation for objects with streaming. We find that our simulations are consistent with some observations at high redshift, but on a population level, they continue to under-predict star formation relative to the majority of observations. Notably, our simulations do not include feedback, and so can be taken as an upper limit on the star formation rate, exacerbating these differences. However, simulations of overdense regions (both with and without streaming) agree with observations, suggesting a strategy for extracting information about the overdensity and streaming velocity in a given survey volume in future observations.

% Streaming simultaneously significantly enhances star formation within low-mass halos, while allowing the formation of gas-rich objects (SIGOs) and associated star clusters outside of halos. We present a study of the population-level effects of streaming on star formation within both halos and SIGOs in a set of simulations with and without streaming. 
%We analyze the efficiency of star formation in differing mass ranges with and without streaming, as well as studying the effects of streaming on the star formation rate density in differing-mass halos and the Kennicutt-Schmidt relation. We additionally compare these quantities to observations in the local Universe and at high redshift. We also discuss the signature of streaming on the UV luminosity function, and for the first time present population-level data on the formation of star clusters in simulated SIGOs. We find that our simulations are consistent \will{Consider wording} with high-redshift JWST observations, and from this re-affirm the potential of increased statistics from coming JWST observations and next generation observations to use the effects of streaming as a test of $\Lambda$CDM at high redshift.

\end{abstract}


\section{Introduction}

The first stars in the Universe, known as Population III (Pop III) stars, formed at zero metallicity and are thought to have very different properties from the stars we see today \citep{Bromm13}. These metal-poor stars formed from gas clouds cooled primarily via radiative transitions of molecular hydrogen (H$_2$). Through this cooling mechanism, primordial gas clouds could lower their temperatures to $\sim200$ K, corresponding to a Jeans mass of $\sim1000$ M$_\odot$. Sufficiently massive primordial gas clouds could then collapse to form stars \citep{Yoshida+08}. Because H$_2$ cooling is less efficient than high-metallicity cooling at very low temperatures, this limits the minimum halo mass at which early Universe structures can form stars \citep{Haiman+96,Tegmark+97,Abel+02,Bromm+02,Yoshida+03early,Lake+22}. These stars were the primary sources of the first metals in the Universe, and as such were vital to later galaxy formation \citep{Ferrara+00,Madau+01}. Understanding the environments in which these first stars formed, and when they formed, will inform our understanding of these later galaxies.

Upcoming and ongoing observations from JWST may have the potential to observe some of these first stars, or the pair-instability supernovae they may produce \citep{Johnson+10,Whalen+13,Visbal+16,Lake+23}. Indirect signatures of Pop III stars, such as the $21$-cm line, also provide a promising means by which to explore Cosmic Dawn \citep{Mebane+20,Magg+22,2023ApJ...958L...3H}. Theoretical models allow us to interpret these observations and maximize their benefit, so there is an immediate need for detailed theoretical models of Pop III star formation \citep{2024arXiv240500813M}. 

% One of the most important parameters in semi-analytic models of Pop III star formation is the minimum mass a halo must have in order to form stars, M$_{\rm crit}$. This parameter is essential for connecting large-scale effects that may be simulated, such as models for halo growth with redshift, to small-scale analytic approximations, such as models of star formation. However, the value of M$_{\rm crit}$ depends on environmental effects, and so in order to make accurate predictions with these models, it is important to understand the ways in which M$_{\rm crit}$ can evolve.

A key environmental effect affecting the formation of Pop III stars is the relative motion of baryons and dark matter in the early Universe, known as the streaming velocity \citep{TH,naoz+13,Kulkarni+21,Hegde+23}. Prior to Recombination, while the growth of baryon overdensities was suppressed by the field of photons, dark matter exhibited bulk flows towards large scale overdensities. These flows were coherent on few-Mpc scales, and their magnitude varied following a Maxwell-Boltzmann distribution. At the time of Recombination, as the temperature of the baryons dropped precipitously, these bulk flows (around v$_{\rm bc, rms} = 30$ km s$^{-1}$ at z$=1100$) became highly supersonic \citep{TH,tseliakhovich11}. In regions with high values of this streaming velocity, gas is capable of advecting from its parent halo, and even when accreted, may exhibit lower densities and star formation rates \citep{TH,naoznarayan14,Williams+22}. Low-mass (M$_{\rm DM} < 10^8 $M$_\odot$) halos in these regions at high redshifts (also known as Dark Matter + Gas Halos Offset by Streaming, or DM GHOSTs) may have their star formation suppressed, as well as exhibiting unique morphological and rotational properties \citep{Williams+22}. As higher values of this streaming velocity are correlated with larger overdensities, this will preferentially impact galaxies and galaxy cluster progenitors at high redshift.

A further impact of this streaming velocity at high redshift is the tendency for gas overdensities to advect out of their parent halos, allowing the formation of gas-enriched objects known as Supersonically Induced Gas Objects, or SIGOs \citep{naoznarayan14,popa,chiou18,Lake+21}. These SIGOs are capable of star formation outside of dark matter halos \citep{chiou+19,Chiou+21,Nakazato+22,Lake+22,Lake+23}. They also form in great abundance in the early Universe, with SIGO abundances approaching the abundance of present-day low-metallicity globular clusters by the epoch of Reionization \citep{Lake+21}. As these SIGOs are composed of nearly pristine gas, this could be an additional source of Pop III star formation prior to Reionization \citep{Lake+22,Lake+23}. 

In this paper, we aim to study the population-level effects of streaming on halos and SIGOs by examining the Kennicutt-Schmidt relation and star formation rate densities in a pair of small-box simulations with and without streaming. The paper is organized as follows: in Section~\ref{sec:Methods} we discuss the setups of the simulations used. In Section~\ref{Sec:Halos}, we discuss the effects of streaming on DM halo star formation, including the Kennicutt-Schmidt relation. In Section~\ref{Sec:SIGOs} we show and contextualize population-level statistics on star formation in SIGOs. Lastly, in Section~\ref{Sec:Discussion} we summarize our results and suggest future avenues of exploration for star formation under the influence of streaming.

For this work, we have assumed a $\Lambda$CDM cosmology with $\Omega_{\rm \Lambda} = 0.73$, $\Omega_{\rm M} = 0.27$, $\Omega_{\rm B} = 0.044$, $\sigma_8  = 1.7$, and $h = 0.71$.

\section{Methodology}\label{sec:Methods}

In this paper, we present the results of two {\tt AREPO} \citep{springel10} simulations, each with $768^3$ DM particles with mass M$_{\rm DM} = 1.1 \times 10^3 $ M$_\odot$, and $768^3$ Voronoi mesh cells with gas mass M$_{\rm B} = 200 $M$_\odot$. Gas cells become eligible to form stars when their mass exceeds the Jeans mass on the cell's scale. Eligible gas cells are converted into star particles on the free-fall timescale. When the free-fall timescale is longer than the simulation timestep, this is implemented as a stochastic process as described in \citet{Marinacci+19}. The formed star particles have the mass of the gas cell that gave rise to them, and are collisionless.

The simulations have a $2.5$ Mpc box size, and are evolved from $z=200$ to $z=12$. Initial conditions are generated using transfer functions from a modified version of CMBFAST \citep{seljak96}, which incorporates first-order scale-dependent temperature fluctuations \citep{NB} and the streaming velocity. In line with the methods of \citet{chiou+19,Chiou+21, Lake+21, Nakazato+22, Lake+22}, we use $\sigma_8 = 1.7$ to generate our initial conditions, simulating a rare overdense region where structure forms early, similar to regions that form galaxy clusters. This enhances our statistical power. One simulation uses a $2\sigma_{\rm vbc}$ = 11.8 km s$^{-1}$ streaming velocity at the initial redshift $z=200$, applied as a uniform boost to the x velocity of the baryons, as in \citet{popa}. The other simulation does not include a streaming velocity effect.

Both simulations explicitly include non-equilibrium molecular hydrogen chemistry and its associated radiative cooling, using the chemistry and cooling library GRACKLE \citep{Smith+17, Chiaki+19}. This includes molecular hydrogen and HD cooling, as well as chemistry for 15 primordial species: e$^-$, H, H$^+$, He, He$^+$, He$^{++}$, H$^-$, H$_2$, H$_2^+$, D, D$^+$, HD, HeH$^+$, D$^-$, and HD$^+$. The cooling rate of molecular hydrogen includes both rotational and vibrational transitions \citep{Chiaki+19}. It is important to note that we do not include metal cooling, which enhances cooling, especially in larger halos that continue to form stars after supernovae. We also do not include Lyman-Werner, radiative, or supernova feedback, which lowers star formation rates by order of magnitude, especially in low-mass halos \citep[see e.g.,][]{Xu+16}. Thus, halos smaller than $\sim10^8$-$10^9$~M$_\odot$ may undergo one burst of star formation. Notably, the highest halo mass in our simulations (few~$\times10^9$~M$_\odot$ ) indicates the overall trend we expect. We highlight that the comparison to the observations is done at the high-mass end.

We use the object classifications from \citet{chiou18} to identify SIGOs and DM halos. We use a friends-of-friends (FOF) algorithm with a linking length that is 20\% of the mean DM particle separation, or about 650 cpc, to identify DM halos. This gives us the locations and virial radii of DM halos in the simulation, assuming sphericity for simplicity \citep[although it is important to note that DM halos at these times can be ellipsoidal e.g.,][]{Sheth+01, Lithwick+11, Vogelsberger+11, Schneider+12, Vogelsberger+20}. We also run the same FOF algorithm on the gas component of the output, with stars as a secondary component. This identifies gas-primary objects. In order for these objects to be considered as possible SIGOs, we additionally require that they contain at least $100$ combined gas and star particles \citep{Chiou+21}.

Following \citep{popa}, we fit gas-primary objects to an ellipsoid, by calculating the smallest ellipsoidal surface that encloses every particle in the object. This is necessary because these gas objects are generally quite elongated within highly non-spherical gas streams. We tighten the ellipsoids by shrinking their axes by 5\% until either the ratio of the axes lengths of the tightened ellipsoid to that of the original ellipsoid is greater than the ratio of the number of gas cells contained in each, or 20\% of their particles have been removed. Finally, in order to be identified as a SIGO we require that the center of mass of these ellipsoids must be located outside the virial radius of nearby DM halos, and that the object must have a baryon fraction above 60\%. These constraints are necessary to effectively distinguish SIGOs from other classes of gas objects \citep{Nakazato+22,Lake+22}.

Finally, because there is some degeneracy between true star-forming SIGOs and errors in the FOF algorithm, we verify each star-forming SIGO visually. The gas and DM density fields around each candidate SIGO are visualized side-by-side, and we visually verify that the gas overdensity associated with each SIGO is spatially offset from nearby DM overdensities. We also visualize the same spatial region from the run without streaming, ensuring that the star-forming SIGO is not present in that (control) run. This gives us confidence that we are identifying true star-forming SIGOs.

We classify all non-SIGO objects that form in our simulations with streaming as DM GHOSts. Corrections for the enhanced $\sigma_8$ in halos both with and without streaming for our star formation rate density plots are made via analytic calculations of the abundance of halos at various masses and redshifts using the methods described in \citet{Lake+21}. These analytic calculations are subsequently used to estimate a redshift-, streaming-, and mass- varying abundance correction to $\sigma_8 = 0.826$, which is applied to our SFR densities. Gas objects are matched to nearby host halos, and these halos' masses are used in the calculations. 

In order to identify stellar components of halos, we run a gas-and-star-primary FOF, and match objects from it to their nearest halos. This allows us to compute stellar masses of halos by summing over associated objects, ensuring that we are identifying even star particles that are just outside their parent halos. In all objects, star formation rates are determined by the formation time of star particles in the simulations. Each new star particle is matched to its host object at the first snapshot after it forms. The implied mass of new stars formed in a given object is divided by the time between snapshots to determine a star formation rate. This represents an average SFR over the timeframe between snaps.

\section{Star Formation in DM GHOSts}\label{Sec:Halos}

\begin{figure*}[t]

\centering
   \gridline{
\fig{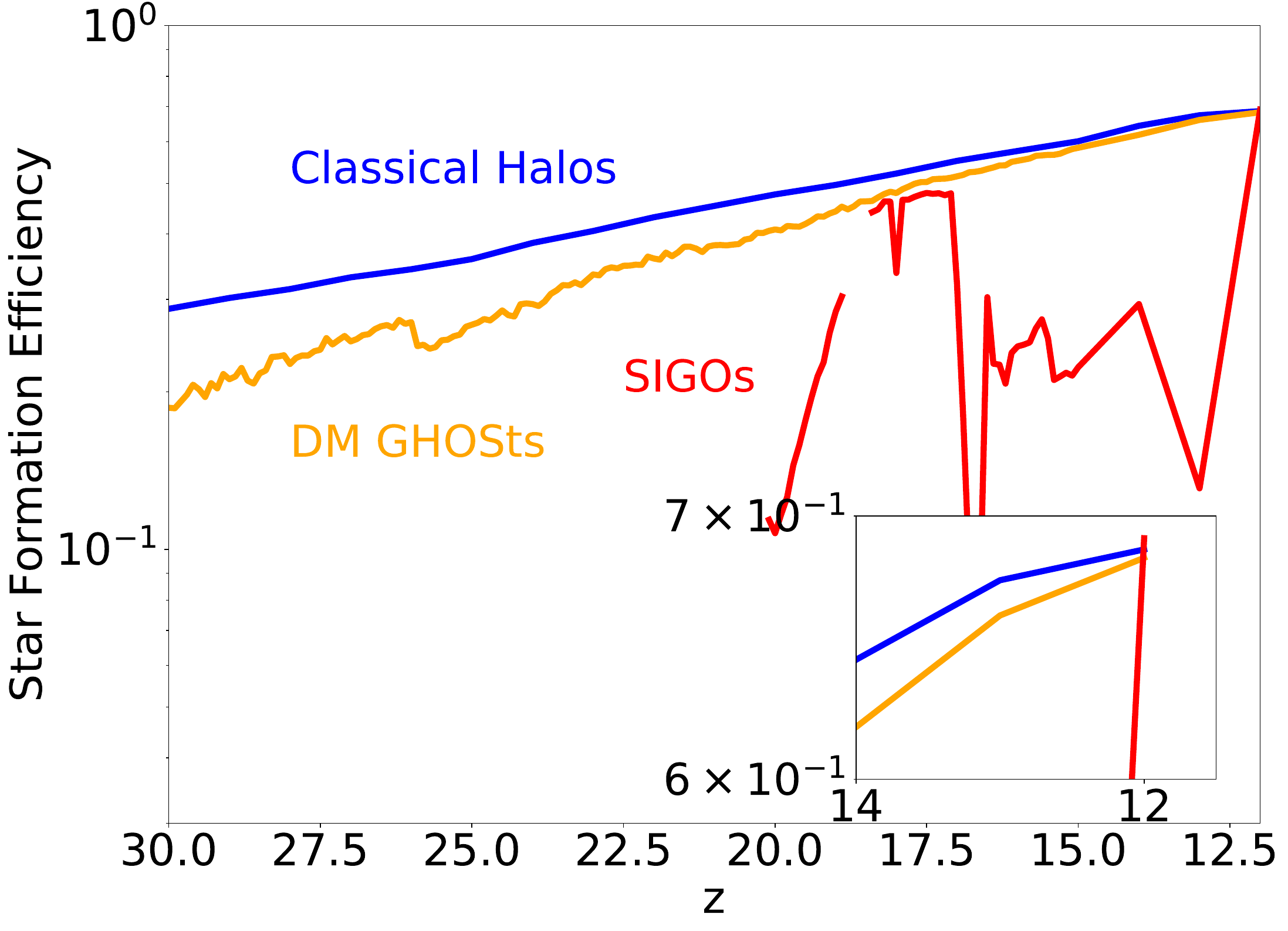}{0.47\textwidth}{\centering Star formation efficiency binned by redshift}
    \fig{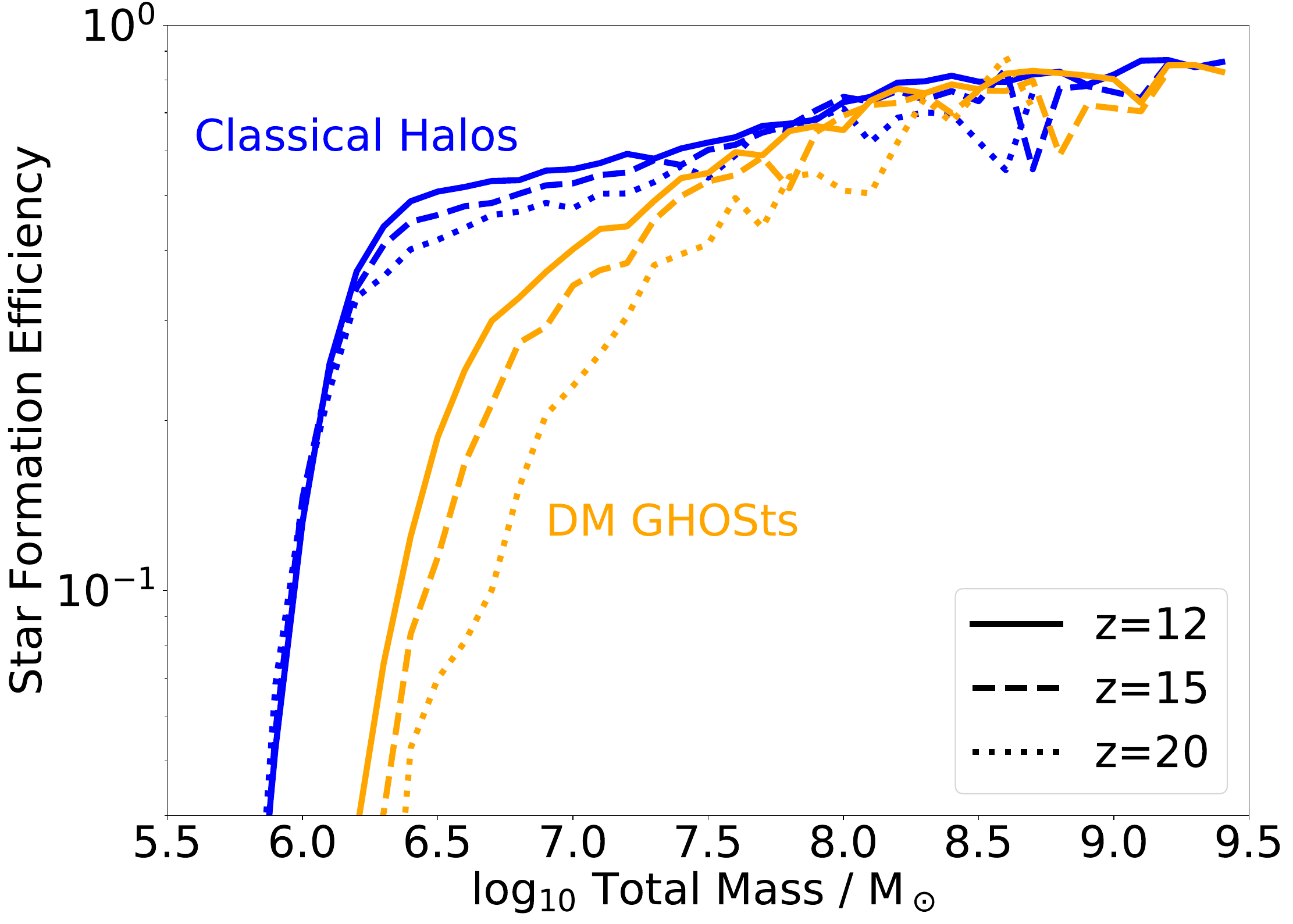}{0.47\textwidth}{\centering Star formation efficiency binned by mass}}
    
    \caption{{\bf \centering Star formation efficiency in various object classes:} here, we show the star formation efficiency of classical halos (halos in our no-streaming simulation, blue), DM GHOSts (halos in our $2\sigma_{\rm vbc}$ streaming simulation, orange), and in the left panel, SIGOs (in our $2\sigma_{\rm vbc}$ streaming simulation, red) as a function of redshift. In the right panel, we show the efficiency of star formation in classical halos (blue) and DM GHOSts (orange) at $z=12$ (solid lines), $z=15$ (dashed lines), and $z=20$ (dotted lines) as a function of mass, binned by 0.1 segments in log$_{10}$ of total mass of all types of matter. At high redshifts, DM GHOSts have substantially suppressed star formation compared to classical halos, but they catch up to the overall star formation rates in halos by $z=12$ (owing in part to their suppression of the abundance of small-scale structure in comparison to larger halos). SIGOs have generally lower overall star formation efficiencies, likely owing to their lack of dark matter.}
    
    \label{Fig:SFE}
\end{figure*}

% \begin{figure}[t]

% \centering
    
%    \fig{SFEMassBinned.pdf}{0.47\textwidth}{}

%     \caption{{\bf \centering Star formation in halos as a function of mass:} Here we show the efficiency of star formation in classical halos (blue, no streaming) and DM GHOSts (orange, $2\sigma_{\rm vbc}$ streaming) at $z=12$ (solid lines), $z=15$ (dashed lines), and $z=20$ (dotted lines), binned by .1 segments in log$_{10}$ total mass of all types of matter. At low masses, DM GHOSts exhibit substantially reduced star formation efficiencies compared to classical halos, but this effect slowly fades and moves to lower mass ranges as redshift decreases. }
    
%     \label{Fig:SFEMassBinned}
% \end{figure}

\begin{figure*}[t]

\centering

    \fig{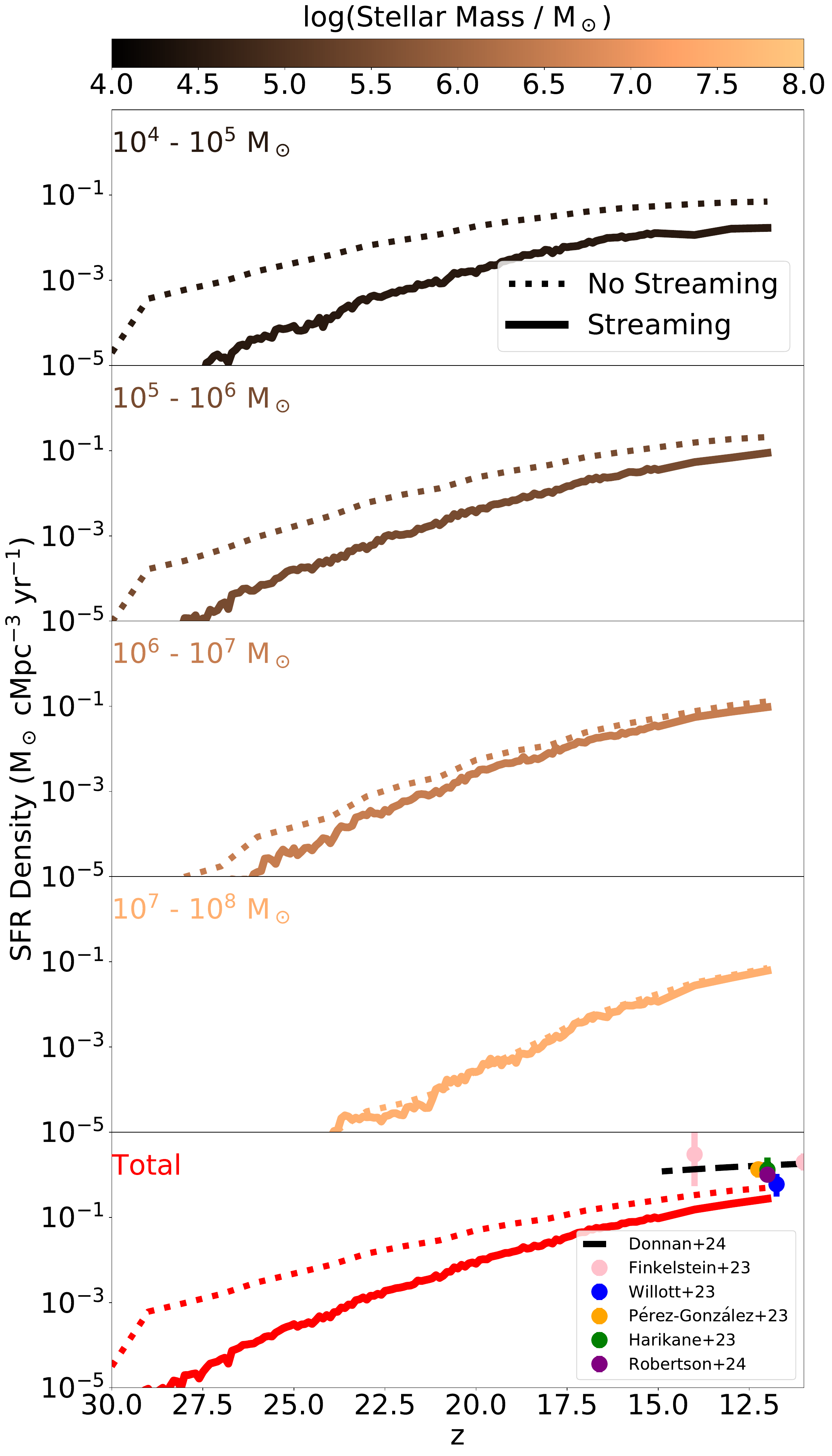}{0.47\textwidth}{Summed SFR Density in Mass Bin}
    \fig{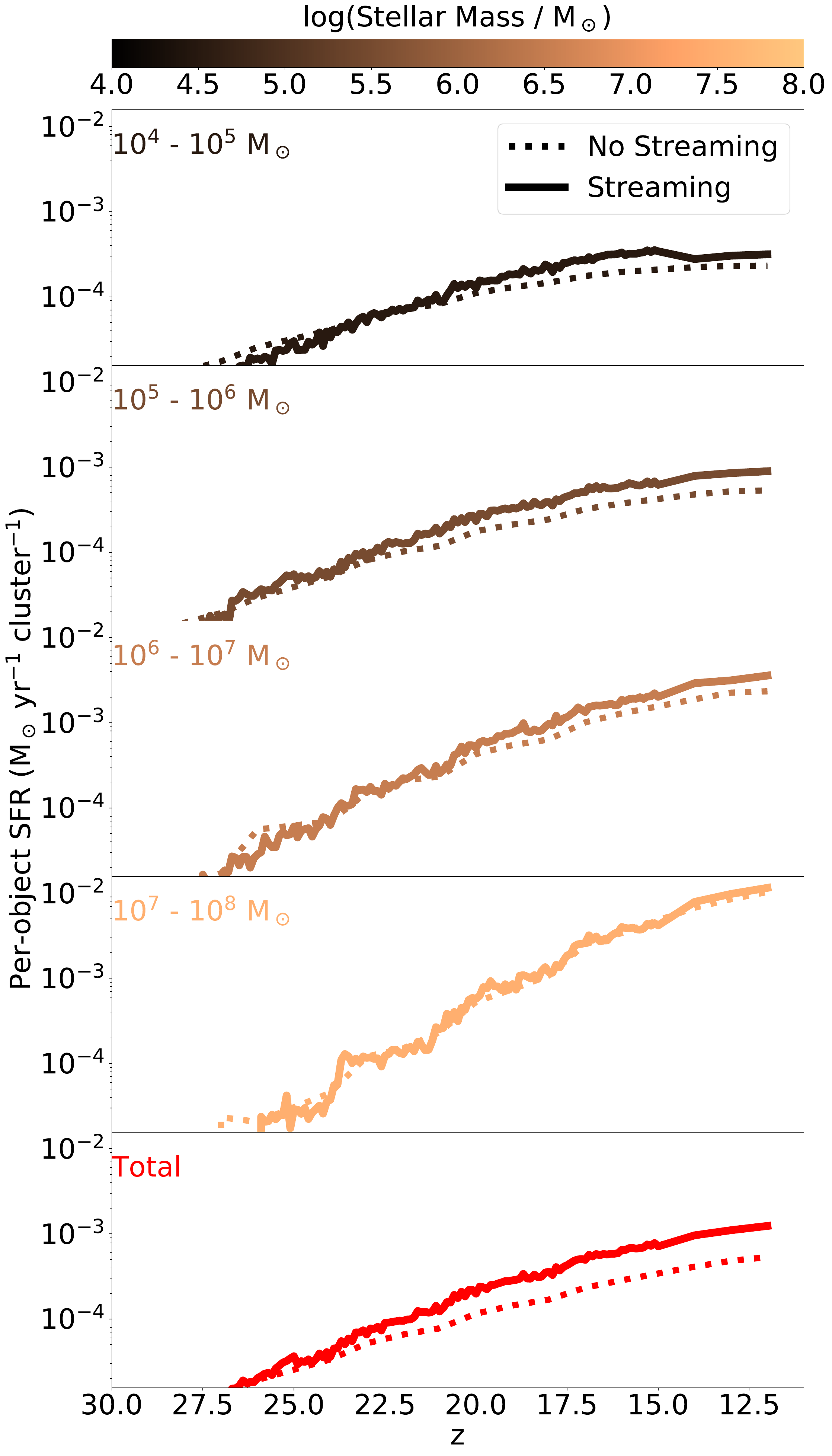}{0.47\textwidth}{Average SFR Density in Mass Bin}
    
    \caption{{\bf \centering Star formation rate density in stellar objects of different masses:} Here we show plots comparing star formation rate densities with (solid lines) and without (dotted lines) streaming at different object masses. The left panel shows population-level statistics and the right panel shows averaged (per-object) statistics. In the left panel, only low-mass object SFRs are substantially lowered by streaming at the lowest redshifts studied on the population level, up to our low-mass resolution limit of about $10^5$~M$_\odot$. Observational constraints from \citet{Donnan+24},  \citet{Finkelstein+23}, \citet{Willott+23}, \citet{PG+23}, \citet{Harikane+23}, and \citet{Robertson+24} are included for comparison in the bottom panel of the left figure, which shows the total SFR density in all objects up to the highest-mass object in our box, representing halo masses from $10^5$~M$_\odot$ up to the highest-mass halo in our box, $2\times10^9$~M$_\odot$. In the right panel, we see that at later times, star formation is actually more rapid in objects of a given stellar mass with streaming than without, as the objects catch up to their no-streaming counterparts. Note that the SFR curves in this Figure were normalized to represent  $\sigma_8=0.826$ using Equation~(\ref{Eq:SigmaCorrection}). The comparison to the observed SFR for the $\sigma_8=1.7$, used in the simulation in shown in Figure~\ref{fig:MadauUnadjusted}. We draw the attention of the reader to the striking agreement between the observations and simulation in the higher $\sigma_8$ case. See Figure~\ref{Fig:sigma8} for further analysis. }
    
    \label{Fig:Madau}
\end{figure*}

\subsection{Star formation efficiency}
DM GHOSts are the counterparts to classical halos under significant streaming velocities and potentially are the ancestors of the oldest dwarf galaxies in our Local Group \citep[][]{Williams+22,Williams+23}. In particular, using observations of nearby dwarf galaxies, it has been pointed out that the local environment probably represented a high streaming patch of the Universe \citep{Uysal+22}. DM GHOSts comprise a stellar mass range that extends far beyond that of SIGOs and, therefore, are readily detectable by today's JWST. Predictions of their star formation history may soon be directly testable. The common prediction in the literature suggests that the streaming velocity suppresses the star formation in DM GHOSts, for DM halo masses $\lsim 10^{8-9}$~M$_\odot$ \citep[e.g.,][]{TH,fialkov2012,oleary12,bovy13,tanaka13,tanaka14,Park+20,Schauer+23,Hegde+23,Conaboy+23}. 

However, \citet{Williams+23} recently showed an additional, more subtle, effect of streaming on star formation. Although in DM halos with halo masses $\lsim 10^7$~M$_\odot$ the number of stars and the star formation efficiency is suppressed at z$\sim 12$ at the population level, the abundance of stars in larger DM halos is not suppressed. This implies a comparable star formation efficiency and a higher star formation rate in high-mass DM GHOSTs compared to classical structures at z$\sim12$. 

 In Figure~\ref{Fig:SFE}, we depict the star formation efficiency, defined as the mass in stars divided by the baryonic mass in a halo, i.e., $M_\star/M_{\rm baryon}$. The left panel of this figure shows the average efficiency over all structure masses as a function of redshift. As shown, the star formation efficiency in halos at $z\sim30$ is about twice as low in halos formed in a simulation with a  $2\sigma_{\rm vbc}$ streaming velocity than in classical halos. This suppression originates from the lower gas densities within halos resulting from streaming at high redshifts, which subsequently suppresses the formation of molecular hydrogen, and thus cooling in low-mass halos. 

In the left panel of Figure~\ref{Fig:SFE}, the star formation efficiency is comparable at $z=12$ in DM GHOSts to that of classical halos \citep[consistent with][]{Williams+23}. To explain this further, in the right panel of this figure we show the star formation efficiency in DM GHOSts (orange) and classical halos (blue) at $z=20$ (dotted lines), $z=15$ (dashed lines) and $z=12$ (solid lines). In the simulation with streaming, the streaming velocity preferentially suppresses the formation of low-mass structures \citep[as expected from analytical work by e.g.][]{TH}. As a result, higher-mass halos, which have higher star formation efficiencies and, with streaming, may have higher star formation rates \citep[e.g.][]{Kravtsov+18,Williams+23}, hold a higher weight in the overall star formation efficiency, leading to comparable or even potentially higher star formation efficiencies as a function of the overall mass of gas in halos. These results are also sensitive to feedback, which is not included here but preferentially suppresses star formation in low-mass halos, as well as significantly lowering the overall star formation efficiency. This feedback is likely to significantly affect the minimum mass cutoff in Figure~\ref{Fig:SFE} by $z=12$ and later redshifts, but in analytic modelling is still important at higher redshifts such as $z=20$ \citep{Hegde+23}.

\subsection{Star formation rate}\label{ssec:Madau}

One of the key results from recent JWST observations \citep[e.g.,][]{Harikane+23,PG+23,Willott+23,Finkelstein+23,Donnan+24} is the presence of unexpectedly high star formation rates in the early Universe. The differing slopes of the star formation efficiencies with and without streaming in Figure~\ref{Fig:SFE} hint that streaming may increase the star formation rate at intermediate redshifts, but the bias towards high-mass objects with streaming in this plot serves as a confounding factor. 

In order to better understand this effect, we turn to population-level star formation rate density statistics in Figure~\ref{Fig:Madau}. This figure bins gas objects by their stellar mass (into factor-of-ten solar mass bins), corrects for our enhanced $\sigma_8$ in our simulations (for comparisons to an average-density volume of the Universe; see Appendix~\ref{sec:appendixSigma8} for more details), and plots their overall comoving star formation rate density across cosmic time. The left panel of this figure shows summed SFR densities in the mass bin, while the right panel shows averaged (per-object) SFR densities. At later redshifts (after $z\sim20$), the population of objects in or above the $10^6$ M$_\odot$ bin have comparable star formation rates with and without streaming. In fact, objects in the $10^7$~M$_\odot$ bin actually have slightly higher star formation rates with streaming at some redshifts, a fact that may derive from enhanced gas accretion \citep{Williams+23}. Furthermore, as seen in the right panel, objects in all mass bins exhibited higher star formation rates individually with streaming than without at lower redshifts, starting at $z\sim20$. This may, again, partially result from enhanced gas accretion. In addition, because star formation is delayed with streaming, objects in a given stellar mass bin here tend to be associated with higher-mass halos with streaming than without it: the higher halo masses with streaming could lead to higher star formation rates for a given stellar mass.

\begin{figure*}[t]

\centering

   \fig{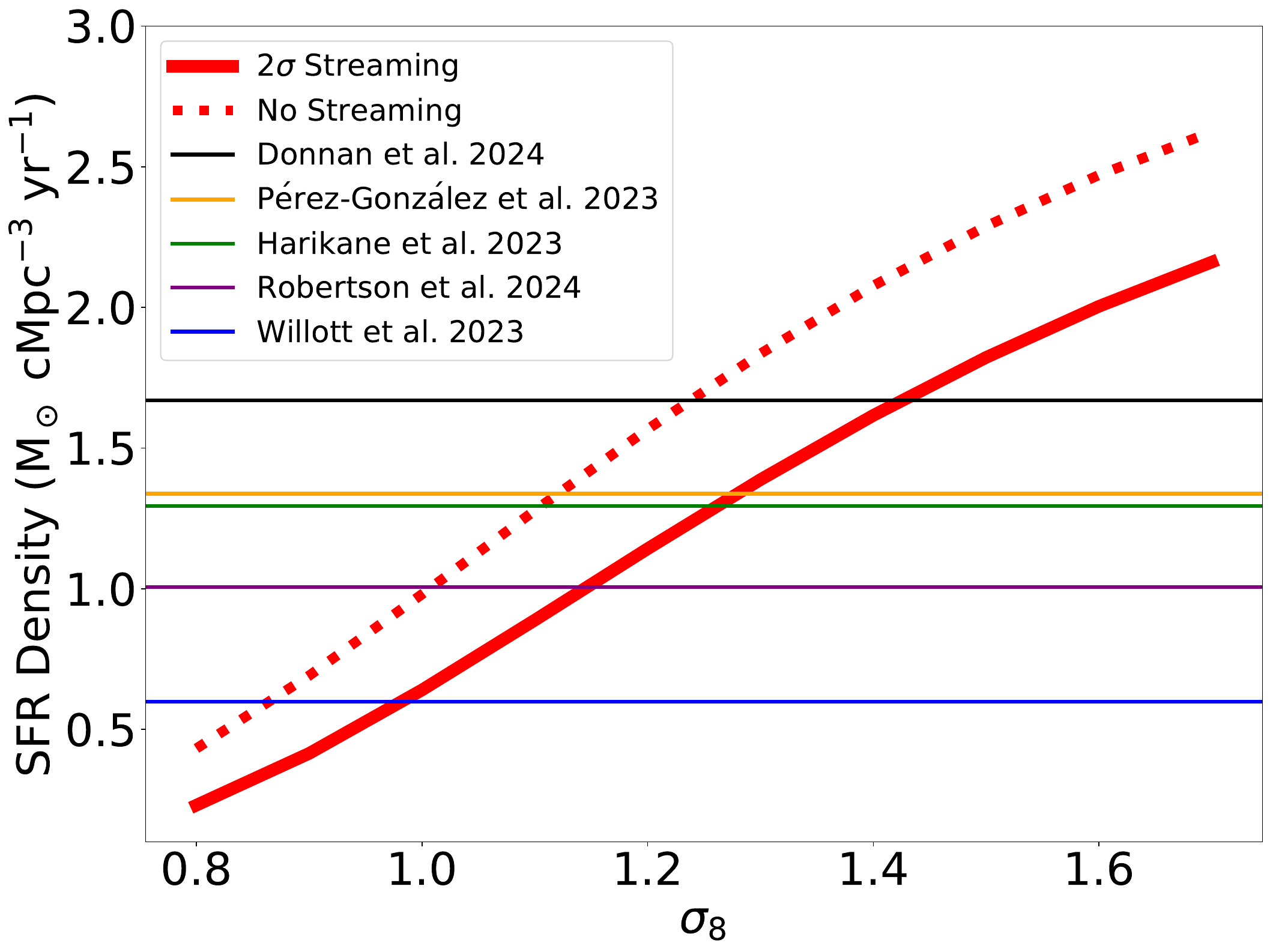}{0.48\textwidth}{}
    \fig{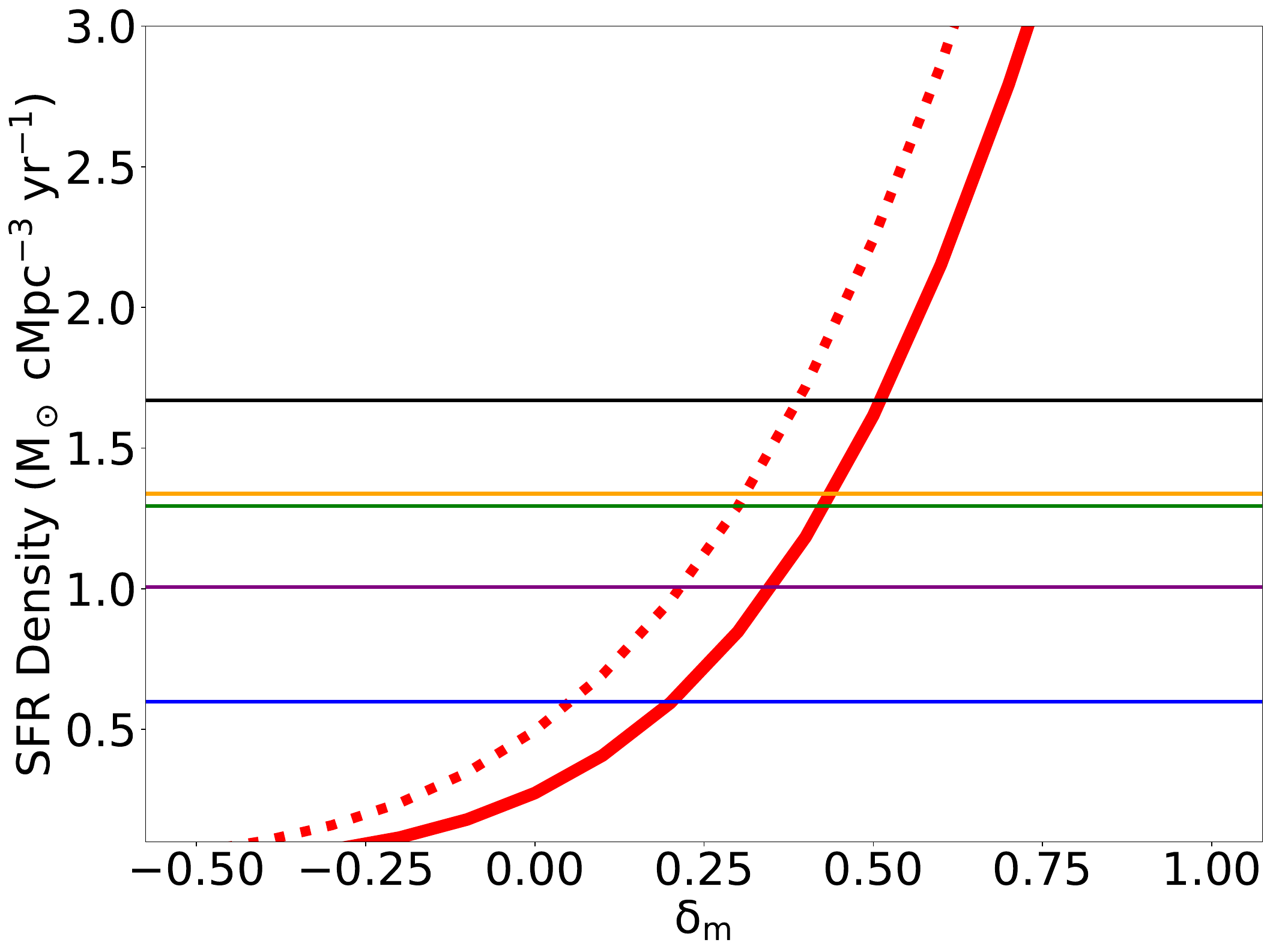}{0.48\textwidth}{}
    
    \caption{{\bf \centering Difference between our simulations and various observations as an overdensity effect:} In the this figure, we show the dependence of our simulations' star formation rate densities at $z=12$ on $\sigma_8$ and the matter overdensity $\delta_{\rm m}$, applying the corrections for varying $\sigma_8$ and $\delta_{\rm m}$ described in Appendix~\ref{sec:appendixSigma8}. The left panel assumes $\delta_{\rm m} = 0$ and the right panel assumes $\sigma_8 = 0.826$. This shows the effect of observing more- or less- dense regions on the predicted star formation rate. Equivalently, this shows a model for the properties of a region of the Universe expected to produce a given observation in these simulations, with and without streaming. To avoid clutter, we've omitted uncertainties, as at present they are large compared to the size of the y axis. Note that this is linearly scaled, while Figure~\ref{Fig:Madau} shows the same quantity but is log-scaled.
    }
    \label{Fig:sigma8}
\end{figure*}

Despite this, the suppression of low-mass halo star formation efficiency is still significant, even at $z=12$ (though, again, at lower redshifts, the mass at which this is significant shifts to lower values). Under the influence of a $2\sigma_{\rm vbc}$ streaming velocity, star formation efficiencies are suppressed by a factor of $\sim2$ or more at halo masses below $\sim10^{7.5}$~M$_\odot$ at $z=12$, as seen in the right panel of Figure~\ref{Fig:SFE}. At higher redshifts, as seen in Figure~\ref{Fig:Madau}, this gap only grows. This may act in concert with the reduced abundances of low-mass halos \citep{TH,Williams+23} to even further suppress the abundance of low-luminosity dwarf galaxies at these redshifts. As the star formation density is constrained to higher redshifts, it is more and more strongly affected by streaming: future JWST and next generation observations may start to constrain this quantity tightly enough at high enough redshift for this to be a good probe of the effects of streaming.

In the left panel of Figure~\ref{Fig:Madau}, we also over-plot several current JWST observational constraints to compare the results of our simulations to the literature. Our results, which are normalized to $\sigma_8=0.826$, are consistent with \citet{Willott+23} as shown, but under-predict star formation relative to the majority of observational constraints. This may be
attributed to reduced molecular hydrogen cooling in our simulation, a resolution effect discussed later, and the lack of metal line cooling. 

However, as we highlight in Figure \ref{Fig:sigma8}, another possibility for this inconsistency is that the observed SFRs are, in fact, coming from an overdense regime \citep[see for an example discussion regarding this point][]{Willott+23}. Therefore, in Figure \ref{Fig:sigma8}, we show the full dependence of our $z=12$ SFR density in each simulation (solid which includes streaming, and dashed without streaming), as a function of the matter overdensity, normalized using $\sigma_8$. The matter overdensity $\delta_m$ is defined in terms of the mean density of matter in the Universe $\rho_{\rm m, mean}$ as \begin{equation} \delta_m = \frac{\rho_m}{\rho_{\rm m, mean}} - 1.\end{equation} The $\sigma_8$ and $\delta_{\rm m}$ correction is specified in Appendix~\ref{sec:appendixSigma8}. We also over plot the star formation rate density observations from Figure~\ref{Fig:Madau} at $z\sim12$, allowing us to convert their observed star formation rates to an equivalent $\sigma_8$ or $\delta_{\rm m}$.

As one can see, the results of \citet{Willott+23} are consistent with regions of the Universe with properties relatively similar to the overall Universe in our simulation without streaming (with an effective $\sigma_8$ value of $0.86$). However, if these observations are derived from a $2\sigma_{\rm vbc}$ patch of the Universe, they are consistent with an overdense regime. Furthermore, at face value, we infer that the \citet{Harikane+23} and \citet{Donnan+24} observations are consistent with $\sigma_8 \approx $ 1.1 (1.26) and 1.23 (1.42) without (and with $2\sigma_{\rm vbc}$) streaming. \citet{PG+23} and \citet{Robertson+24} have similarly overdense inferred fields with $\sigma_8 \approx $ 1.11 (1.27) and 1.01 (1.14) without (and with $2\sigma_{\rm vbc}$) streaming.

Naturally, other factors, such as metal line cooling or observational corrections, may contribute to the uncertainty of this analysis. Regardless, we suggest a strategy that can either extract the underlining density of the star-forming region or, in case the latter is found via other means, can constrain the value of streaming. Lastly, Figure~\ref{Fig:sigma8} suggests that the aforementioned observations are consistent with overdense regions in the Universe, apart from \citet{Willott+23} which is consistent with a near-mean-density region without streaming.

\begin{figure*}[t]

\centering

   \fig{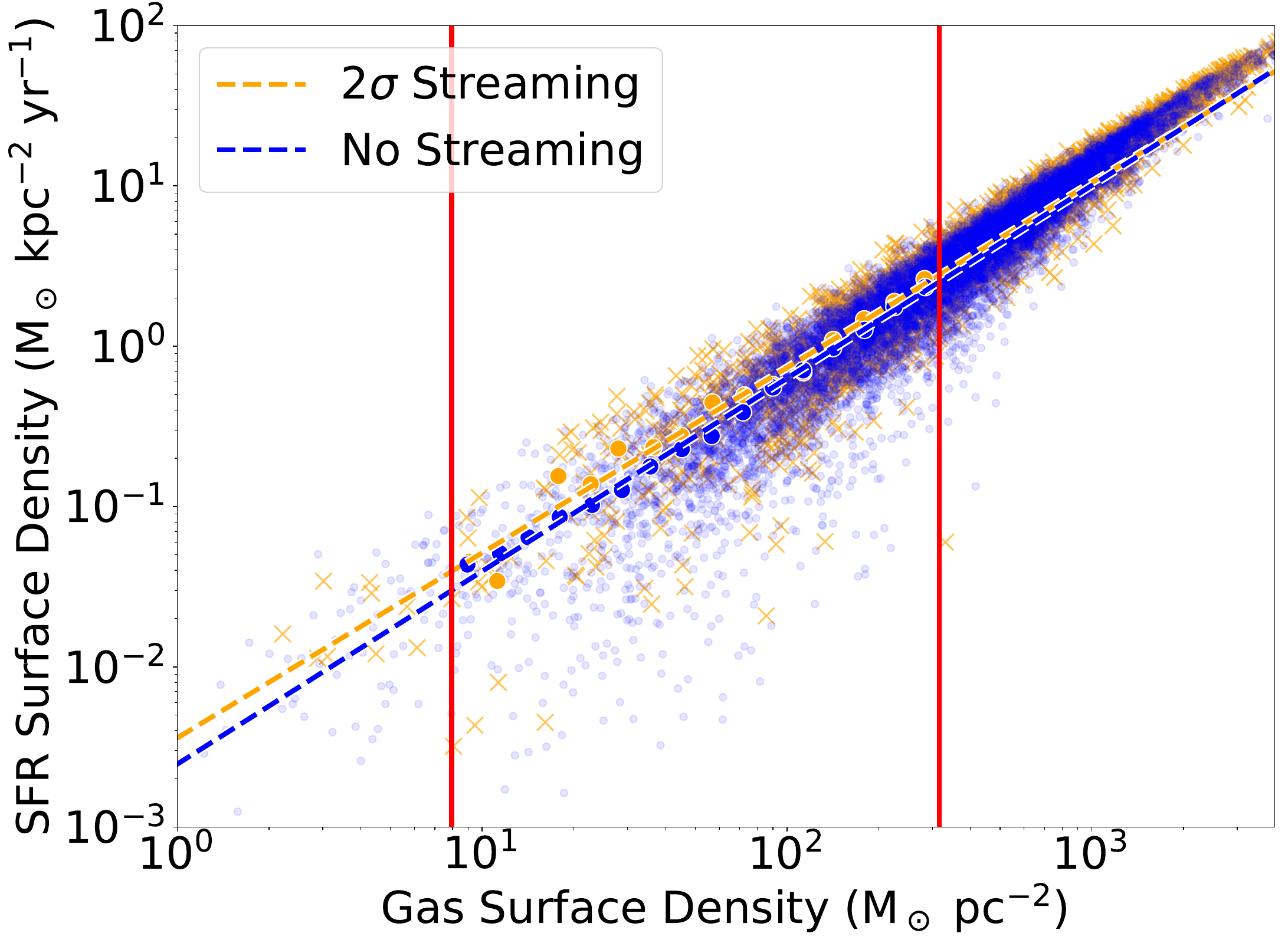}{1.\textwidth}{}

    \caption{{\bf \centering Kennicutt-Schmidt relation (KSR) in DM GHOSts:} In the this figure, we show the relation between the star formation rate surface density and gas surface density in halos with (orange) and without (blue) the influence of baryon-dark matter streaming, at $z=12$. Individual halos' star-forming regions are plotted as individual points, and trend lines are linearly fit and over-plotted with dashed lines of matching colors. Low-gas density halos are suppressed by streaming, necessitating a low-density cutoff to minimize sampling bias. Furthermore, the highest-density halos are impacted by our time resolution, with a KSR slope that approaches $1$ as the minimum density considered is raised. However, when comparing only moderate-density halos (region between the red lines), the KSR has consistent slopes with and without streaming. The range of our fit is shown as a red-outlined central region.}
    
    \label{Fig:KennicuttSchmidt}
\end{figure*}

\subsection{Kennicutt-Schmidt relation}

A priori, the aforementioned result of enhanced star formation for a given object seems counterintuitive, given the injection of turbulence associated with streaming. However, the DM halo provides a deep potential well for the gas in the DM GHOSts, which overcomes the gas pressure. This effect is then reflected in the dependency of the SFR with the gas surface density, otherwise known as the Kennicutt-Schmidt relation \citep[KSR,][]{Schmidt59}. In Figure~\ref{Fig:KennicuttSchmidt}, we compare the KSR within DM GHOSts in our $2\sigma_{\rm vbc}$ streaming run (orange crosses) to that in our run without streaming (blue dots) at $z=12$. Gas surface densities are taken within an ellipsoid described by the smallest ellipsoid capable of containing all of a given gas object's star particles. In this proof of concept, the SFR surface densities considered are those of star particles formed between $z=13$ and $z=12$.
Halos whose star-forming regions convert more than $25\%$ of their gas mass to stars in this timespan are omitted, as we do not have the time resolution to adequately trace their star formation as a function of a defined, constant gas density.

As one would expect, we see a trend of increasing SFR density with gas surface density with and without streaming. 
To differentiate between the SFRs of the two cases at given gas densities, we use linear regression to subtract the rising trend in the data. As mentioned, one effect of streaming on these objects is to prevent low-mass/low-density objects from forming stars at these redshifts: this creates a bias towards low-density behavior in our run without streaming, which could be different from intermediate-density behavior. Furthermore, our limited time resolution and lack of feedback cause high-density objects to tend towards a slope of $1$, representing the full conversion of gas to stars. In an effort to lessen the impact of missing feedback, we address these issues by introducing high- and low-density cutoffs in the data (see Appendix~\ref{sec:appendix} for more details), and bin the remaining data into bins of 0.1 dex width in gas surface density space. Gas surface density values and SFR surface density values are calculated from the mean of the objects in each bin, and SFR surface density errors are taken as the error of this mean. To account for uncertainties derived from the binning process, the gas surface density errors are taken to be half the width of the bin or 0.05 dex. Using linear regression, we then find a best-fit slope $\alpha$ in each case that fits the relation 
\begin{equation}
    {\rm log}\left(\Sigma_{\rm SFR}\right) = \alpha\times{\rm log}\left(\Sigma_{\rm gas}\right) + C \ .
\end{equation}\label{eq:KSR}

In Figure~\ref{Fig:KennicuttSchmidt}, we examine the moderate-density star-forming gas objects left after the cuts and show our fit for these objects. Our high- and low-density cuts are shown as red lines. We overlay lines of best fit for the KSR in our $2\sigma_{\rm vbc}$ ($0\sigma_{\rm vbc}$) streaming case in blue (orange). These represent $\alpha = 1.20\pm0.02$ for the case without streaming and $\alpha = 1.16\pm0.05$ for the case with $2\sigma_{\rm vbc}$ streaming: the difference between these two slopes is merely the standard error. In fact, the only noticeable difference is that the moderate-density star formation regions with streaming may have somewhat higher star formation rates for a given gas density. With the slope estimates subtracted and marginalizing over the independent errors in both slopes and assuming that the residuals are normally distributed, we compute a probability of $p=0.0012$ that the difference in the means of the residuals with and without streaming is due to random chance. Thus, there is sufficient evidence that the star formation rate density in gas regions of the same density is higher with streaming than without it. While surprising, this can perhaps be explained using the results of \citet{Williams+23}: gas which would have been accreted onto low-mass halos at high redshift was instead advected out in the run with streaming. This gas, which is still generally localized in high-density regions, enhances accretion flows onto halos at lower redshifts, potentially resulting in enhanced star formation in higher-density regions around $z=12$ under the influence of streaming compared to the no streaming case. This manifests itself in stronger star formation rates for a given gas surface density with streaming in this plot.

\begin{figure}[t]

\centering
   
\includegraphics[width=0.5\textwidth]{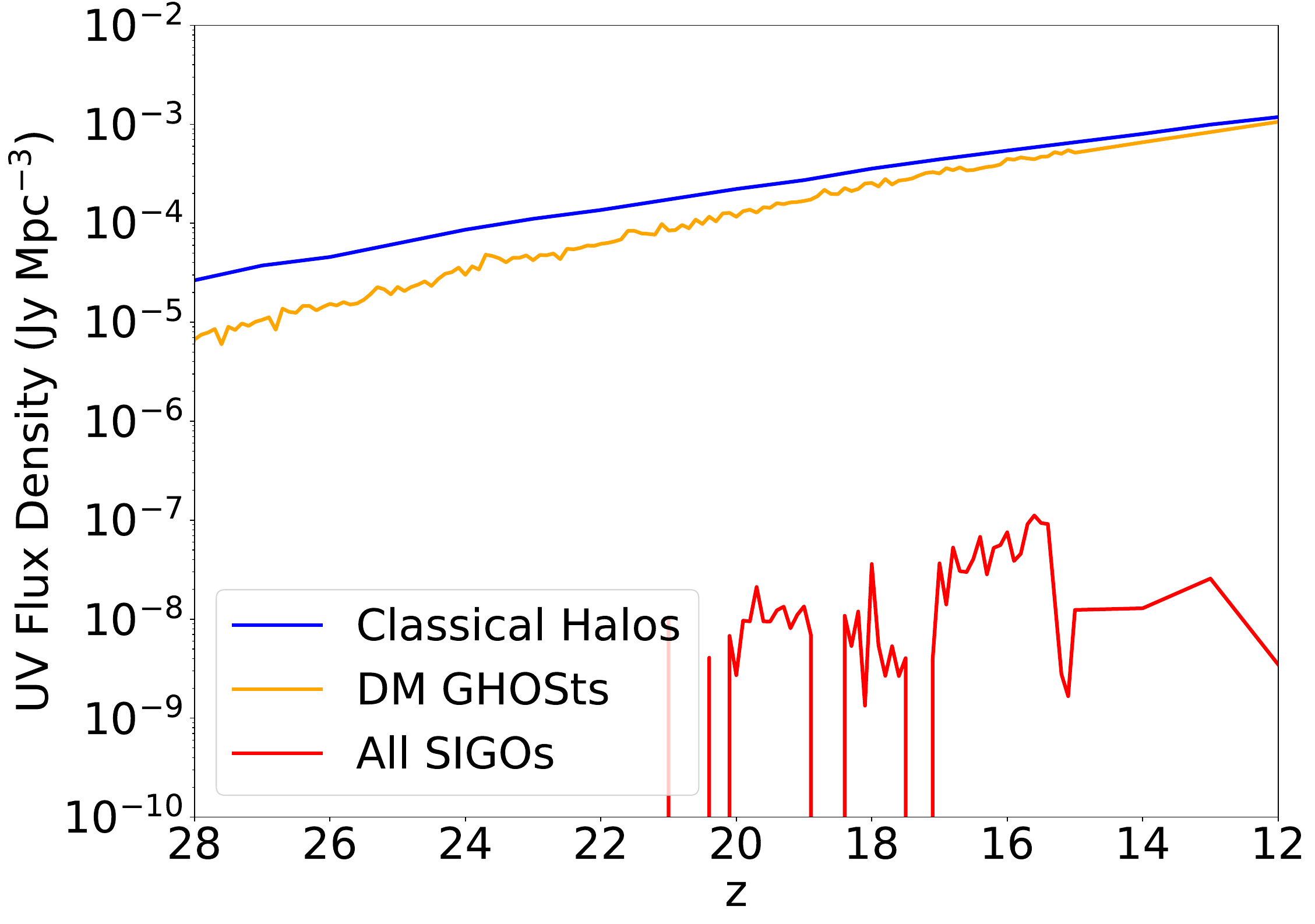} 
    
    \caption{{\bf \centering Contributions of object classes to the UV luminosity function} Here, we show the UV flux density of classical halos (halos without streaming, blue), DM GHOSts (halos with a $2\sigma_{\rm vbc}$ streaming effect, orange), and SIGOs (red, with $2\sigma_{\rm vbc}$ streaming). This relies on new stars--over time. Although the DM GHOSts form stars more slowly at high redshifts, the influence of streaming fades and they begin to form stars as rapidly as classical halos. Meanwhile, SIGOs form stars several orders of magnitude slower, owing to their relatively low masses, and are a small contribution to the overall UV continuum at high redshift.}
    
    \label{Fig:Detectability}
\end{figure}

%Smadar is here

 Most of the flux from these early objects takes the form of UV emission from young, massive stars \citep[e.g.,][]{Sun+16,Hegde+23,Senchyna+23}. We can estimate this flux semi-analytically using the method of \citet{Sun+16}: \begin{equation}\label{Eq:luminosity}
\dot{M}_{\rm SFR} = \mathcal{K}_{\rm UV,1500}\times L_{\rm UV,1500} \ ,
\end{equation}
where we take $\dot{M}_{\rm SFR}$ to be the average star formation rate in a given class of objects within a ${\rm \Delta z} = 0.1$ period preceding the redshift examined (effectively showing the averaged-out flux from a given class of objects on large scales). $L_{\rm UV,1500}$ is the rest-frame UV luminosity at $1500$ \AA. $\mathcal{K}_{\rm UV,1500}$ here is a fiducial constant, which following \citet{Sun+16} we set to be approximately $\mathcal{K}_{\rm UV,1500} = 1.15\times10^{-28}$ ${\rm M}_\odot {\rm yr}^{-1} / {\rm ergs}$ ${\rm s}^{-1} {\rm Hz}^{-1}$, which assumes a Saltpeter IMF. At these early redshifts, this may well underestimate luminosities, as the metallicities of these early objects are likely quite low and could permit a top-heavy mass function.

Figure \ref{Fig:Detectability} shows the total flux density from classical halos and DM GHOSts as a function of redshift, in blue and orange respectively, compared to the flux from SIGOs in red (the SIGOs are discussed in Section~\ref{Sec:SIGOs}). These UV fluxes are calculated with Equation~\ref{Eq:luminosity}. SIGOs are only included as long as they have not merged with halos -- several SIGOs fall into halos over the course of this simulation (analogous to accreted clusters), and once this happens they are included as part of the DM+G object flux for the purposes of this figure.

Under the influence of streaming, UV fluxes from DM GHOSts at very high redshifts ($z>20$) are nearly an order of magnitude lower than UV fluxes from classical halos. However, as with star formation efficiencies, this gap closes over time, and by $z=12$, nearing the era of reionization, this gap has nearly closed completely. This is an even stronger statement of the fading effect of streaming than the narrowing gap in star formation efficiency seen above: here, this result does not depend on the gas mass in halos. Since as seen in Figure~\ref{Fig:sigma8} the star formation rate in these objects is still a sensitive probe of streaming at $z=12$, this implies that future, higher-redshift observations will be even more sensitive.

\section{Star Formation in SIGOs}\label{Sec:SIGOs}

\begin{figure}[t]

\centering
   
\includegraphics[width=0.5\textwidth]{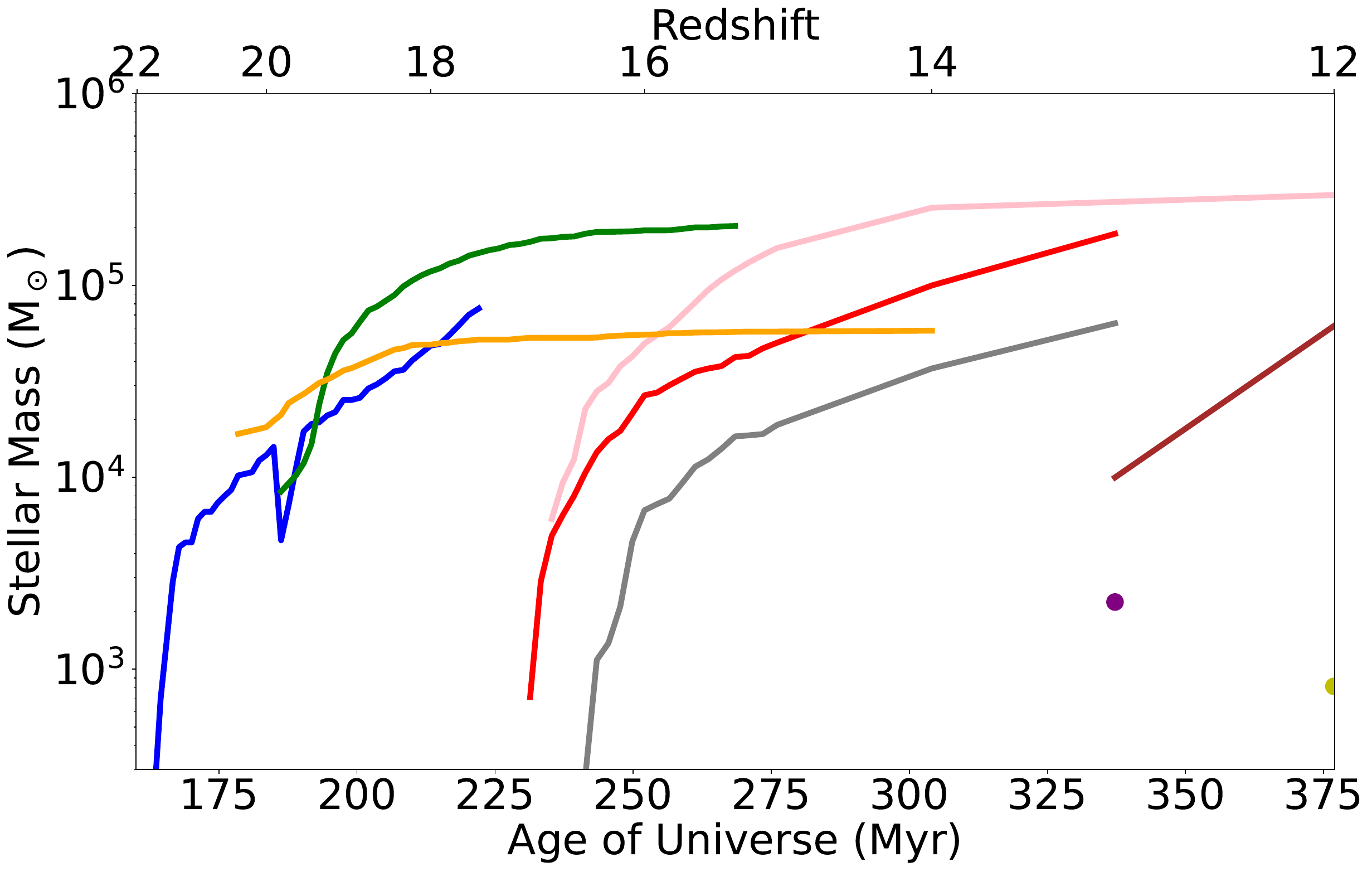} 
    
    \caption{{\bf \centering The star formation history of SIGOs:} This plot shows the star formation history of all visually confirmed SIGOs which form stars in our $2\sigma_{\rm vbc}$ simulation. The first of these SIGOs to form stars does so before $z=21$, and the final SIGO to begin star formation does so at the final timestep of our simulation, $z=12$. SIGOs are removed from this plot when they merge with nearby dark matter halos. Two of these SIGOs (colored in blue and green) fall all the way to the center of the nearest halos, becoming structures akin to nuclear star clusters.}
    
    \label{Fig:MStarvsZ}
\end{figure}

As has been recently shown through both simulations and theory \citep{Lake+22,Lake+23,Nakazato+22}, SIGOs are capable of forming stars, even outside of dark matter halos. These star clusters may be accreted by forming protogalaxies, becoming bound clusters analogous to globular clusters \citep{Lake+23}. In this section, we analyze population-level properties of these clusters, aiming to estimate their mass range and abundances.

In Figure~\ref{Fig:MStarvsZ}, we show the star formation history of each star-forming SIGO in our simulation. Each SIGO is represented by a different-colored line, and (two) SIGOs that are only found with stars at a single snapshot are represented with points. Each of these SIGOs is visually verified by visualizing their gas and dark matter density evolutions, verifying that the gas overdensities that give rise to the SIGOs originate outside of DM halos. 

$9$ SIGOs in this simulation form stars by $z=12$, an abundance of $0.58$~cMpc$^{-3}$. Keeping in mind that these simulations do not include feedback and thus tend to convert all gas to stars on long timescales in star-forming objects, the characteristic masses of stars formed quickly (i.e. on $\sim 10$~Myr timescales) in these objects fall within the range M$_* \sim10^3 - 10^5$~M$_\odot$. These SIGOs are diverse in their evolutionary outcomes: for example, of the $9$ star-forming SIGOs in our simulation, $2$ fall into the centers of dark matter halos and appear to form objects analogous to nuclear star clusters, while $7$ remain visually distinguishable baryon objects by the end of our simulation, $z=12$. 

As many of these SIGOs will disperse through processes such as two-body relaxation before the present day \citep[e.g.,][]{naoznarayan14,Lake+22}, observational campaigns for SIGOs must focus on high redshifts, where these objects are abundant, albeit faint. As shown by \citet{Lake+23}, massive SIGOs may be detectable by current instruments, so it is important to constrain the properties of these objects at high redshift, to aid in distinguishing them from other classes of object and contextualizing observations.

As seen in Figure~\ref{Fig:Detectability}, the UV flux density from SIGOs in the simulation is several orders of magnitude lower than the UV flux from DM+G objects, owing to their lower abundance, and the significantly higher maximum mass of DM+G objects, compared to the maximum stellar mass of SIGOs of about $10^5$~M$_\odot$. As noted by \citet{Lake+23}, SIGOs are not likely to be detectable by current generation instruments at these simulation redshifts ($z\geq12$) owing to these low luminosities. However, similar SIGOs forming later, just preceding Reionization, may be detectable \citep{Lake+23}, making it important to understand the process of star formation within SIGOs.

As SIGOs form outside of dark matter halos and follow a relatively unique evolutionary pathway to form stars as a result, it is interesting to compare the efficiency of their star formation to that of star formation in DM GHOSTs (gas components of halos in regions affected by significant streaming; in this case, halos in our $2\sigma_{\rm vbc}$ streaming velocity simulation) and classical halos (halos without streaming). To this end, we can revisit Figure~\ref{Fig:SFE}, which shows the efficiency of star formation. The gap in the SIGO data in this figure between $z=20$ and $z=19$ corresponds to redshifts where no star-forming SIGOs are found outside of halos (prior star-forming SIGOs have merged into halos).

Owing to small number statistics, SIGOs have the most variability in their star formation efficiencies as a function of redshift, but generally have the smallest star formation efficiencies of the three object classes.  Potentially owing to their lack of dark matter and reliance on molecular hydrogen cooling, SIGOs tend to form stars more slowly and less efficiently than classical halos \citep[e.g.][]{Lake+22}. In addition, in this simulation that does not include radiative feedback, older objects tend to exhibit artificially high star formation efficiencies. Because older SIGOs tend to merge with nearby halos and drop out of this figure \citep[][]{Lake+22}, while older DM+G objects tend to survive and keep forming stars, the calculated efficiencies of star formation in classical halos and DM GHOSts may perhaps be biased to be relatively higher than in our observed Universe compared to SIGOs. This effect will be particularly pronounced for low-mass DM GHOSts.

\section{Conclusions}\label{Sec:Discussion}

The streaming velocity has a variety of significant effects on star formation at high redshifts, from suppressing star formation in low-mass halos to allowing the formation of small-scale gas structures outside of halos and even potentially enhancing star formation at certain redshifts in high-mass halos \citep[e.g.,][]{naoznarayan14,fialkov+14,chiou+19,Schauer+21,Lake+23,Hegde+23,Williams+23}. Notably, it has been suggested that our own Local Group may have formed in a region with a particularly high streaming velocity, increasing the significance of streaming's effects for our understanding of our own galaxy \citep{Uysal+22}. In this work, we focused on two components of these effects: the population-level contribution of supersonically induced gas objects (SIGOs) to early star formation and the impacts of streaming on star formation rates in early halos. 

In order to contextualize the role of the streaming velocity in early star formation, we present the first population-level study of star formation in SIGOs and DM GHOSts in a simulation with explicit star particles. We estimate the stellar masses of these objects and study how they change over time. We systematically compared them to classical objects (i.e., no stream velocity). For example, as seen in Figures~\ref{Fig:SFE} and \ref{Fig:Madau}, significant differences in star formation efficiencies and rates persist between low-mass halos to as late as $z=12$ in these simulations with and without streaming. Streaming tends to advect gas out of low-mass halos, an effect which persists even in $\lsim 10^7$~M$_\odot$ halos at $z=12$. 
Regions of the Universe with relatively high values of the stream velocity, such as our Local Group, may then have critical masses for star formation from the molecular hydrogen cooling threshold that rival those from Lyman-Werner radiation even at these relatively late redshifts \citep[e.g.,][]{Hegde+23}. This is clearly visible in the right panel of Figure~\ref{Fig:SFE}, where the mass at which the star formation efficiency in DM GHOSts drops off is systematically much higher than in classical halos. 
  
At $z\sim 12-20$, accretion from the surrounding gas is enhanced with streaming for high-mass halos, resulting in slightly higher overall SFR densities in halos with streaming than without it after $z=20$. This effect takes place %is true 
even for lower-mass objects which are classically treated as being inhibited by streaming. The SFR per stellar object is higher in the case of streaming in all mass bins at $z=12$, as depicted in the right panel of Figure~\ref{Fig:Madau}. %We note that the SFR 
In Figure~\ref{Fig:KennicuttSchmidt}, we also see that the SFR surface density (i.e., the Kennicutt-Schmidt relation) is enhanced in moderate-gas-density (roughly, lower mass) halos with streaming compared to without it at $z=12$. 

These effects are studied in the case of a $2\sigma_{\rm vbc}$ streaming velocity; however, $1\sigma$ streaming velocities are more common in the Universe, and while rarer, $3\sigma$ streaming velocities are present and present a more extreme case. To infer the differences between these cases and the present case, and with an eye on the accretion-driven origin of this SFR enhancement, we must understand when accretion is enhanced for different values of the streaming velocity. This has previously been studied in the form of the gas fraction in halos \citep[][]{naoz+13, Hirano+23}, which catches up to the no-streaming case at different redshifts depending on the value of the streaming velocity in the region. In the more common lower-streaming cases, the suppression of the gas fraction in halos is weaker and ends at higher redshifts (e.g. $z\sim15$). Therefore, we expect weaker impacts on star formation rates in this case that are more pronounced at higher redshifts. On the flipside, in higher-streaming regions, star formation rates in individual halos could be even more enhanced at $z\sim12$, with effects that persist into the epoch of reionization.

It is important to note that our simulation is resolution-limited and does not include feedback or metal line cooling, which is likely to cause a general lowering and flattening with redshift of the SFR density relations we show, as feedback is more important at lower redshifts. These competing effects likely lead to an underestimation of the efficiency of molecular hydrogen cooling, as well as that of cooling more generally \citep{Nakazato+22}. The lack of feedback also likely leads to an overestimation of the star formation efficiency in these objects, especially at later times and at low masses. This is particularly pronounced in the minimum mass at z=12 and in the low-mass star formation rates at z=12. Thus, we are most sensitive to star formation in halos in these simulations in the DM mass range of a few$\times10^6$~M$_\odot$ or higher, where our results are less sensitive to molecular hydrogen cooling rates \citep[instead, they depend on the filtering mass, e.g.,][]{Hegde+23}. 

% Despite these caveats, our simulations are broadly in agreement with the recent JWST observations. As shown in Figure~\ref{Fig:Madau}, some current observational constraints from JWST are compatible with our simulations without streaming given their margins of error \citep{Willott+23}, although we do under-predict star formation relative to larger surveys such as \citet{Donnan+24}, likely attributable to these resolution effects and the lack of metals in our simulations.

As shown in Figure~\ref{Fig:Madau}, some current observational constraints from JWST may already be compatible with our simulations without streaming, given their margins of error \citep{Willott+23}. However, our SFRs may be considered upper limits due to the lack of feedback, thus exacerbating the discrepancy between simulated and observed star formation rates. In principle, as suggested by Figure 3, comparisons between observed and simulated SFRs can permit inferences on the environments of observed systems, such as their large-scale over- or under- density and their streaming conditions.

We note that the SFR shown in Figure~\ref{Fig:Madau} is normalized to the Planck 2013 $\sigma_8=0.826$ \citep{Planck14}, chosen for consistency with previous studies of DM GHOSts. However, a smaller $\sigma_8$ seems to have similar effects to a decreased large-scale overdensity. We analyze the effect of the dense environment (i.e., the ``local'' $\sigma_8$ or $\delta_{\rm m}$) in Figure \ref{Fig:sigma8}, where the SFR density at $z=12$ for varying large-scale overdensities (see Appendix~\ref{sec:appendixSigma8}), are shown with (solid) and without (dashed) streaming. This Figure highlights the potential of comparing observations with simulations. Specifically, constraints on the over- or under-density of the observational volume and its star formation rate may serve as an indication of the stream velocity. For example, \citet{Willott+23} speculated that their observations may take place in an underdense region. This observation is consistent with our no-streaming simulations. On the other hand, high SFRs such as the one reported by \citet{Donnan+24} may require a $\sigma_8=1.42$ ($\sigma_8=1.23$) with (without) streaming. Once the relation between today's observations and simulations is better understood, a similar strategy can be used to extract information about the density and streaming value of future survey fields.

Finally, the streaming velocity produces SIGOs. Figure~\ref{Fig:MStarvsZ} shows the star formation history of all SIGO-derived star clusters in the simulation. Out of $5325$ unique SIGOs found in the simulation at all redshifts, only $9$ confirmed SIGOs formed stars outside of halos, showing the rarity of this process as a function of the overall abundance of SIGOs (which is not surprising for low-mass objects outside of DM halos). However, as SIGOs are common objects before reionization, this still represents an abundance of SIGO-derived star clusters of $0.58$~cMpc$^{-3}$, which is higher than the local abundance of low-metallicity globular clusters \citep[$0.46$~cMpc$^{-3}$,][]{Rodriguez+15}. Typical stellar masses for SIGOs in this simulation within $10$~Myr after the beginning of star formation\footnote{Feedback will dominate after this period and so we think it is important to emphasize the results of this initial burst of star formation, which has a more limited star formation efficiency.} fall on the order of $10^3 - 10^5$~M$_\odot$; however, it is important to remember that this is likely sensitive to the efficiency of molecular hydrogen cooling, which as noted above is underestimated in our simulation \citep{Lake+22}. It is, therefore, possible that this is an underestimate of the final stellar mass of SIGOs.

Given that SIGOs form stars, it is interesting to ask whether the total stellar flux from SIGOs is significant compared to the flux from classical objects at high redshift. The simple answer is that, owing to their small stellar masses, UV emission from SIGOs is relatively small (Figure~\ref{Fig:Detectability}). However, we suggest that as SIGOs sit within the gas streams which feed early halos, feedback from these SIGOs may have important effects on gas accretion onto these early halos, potentially lowering the star formation efficiency in halos when feedback is considered in conjunction with streaming.

Also visible in Figure~\ref{Fig:Detectability} is a comparison of UV flux from more classical halos with (orange) and without (blue) streaming. An important feature of streaming can be observed here: while there is a near order-of-magnitude dropoff in UV emission with streaming at $z=30$, the effect of streaming falls off with redshift (as the streaming velocity decays), such that the UV emission with and without streaming is similar by $z=12$, as star formation rates have caught up within halos in the streaming velocity simulation. Similarly, for older stars, Figure~\ref{Fig:SFE} shows that the overall star formation efficiencies in classical halos and DM GHOSts have begun to converge by $z=12$. This also results in part (right panel of Figure~\ref{Fig:SFE}) from the increasing contribution of high-mass halo star formation, which is minimally affected by streaming, to the overall rate of star formation in the Universe at lower redshifts.

As this paper shows, observations have, for the first time, begun to directly probe and place constraints on the era where the streaming velocity is critical to structure formation. As the JWST dataset grows and observational constraints at high redshifts tighten, the potential for statistical comparisons at these high redshifts to isolate the effect of streaming grows with it. Because the streaming velocity is a prediction of $\Lambda$CDM not present in many alternative models, this presents an exciting opportunity for coming high-redshift observations to test $\Lambda$CDM structure formation.

\acknowledgments

%The authors thank Steve Furlanetto and Sahil Hegde for helpful conversations. 
W.L., S.N., Y.S.C, B.B., F.M., and M.V. thank the support of NASA ATP grant No. 80NSSC20K0500 (19-ATP19-0020), and the XSEDE AST180056 allocation, as well as the
Simons Foundation Center for Computational Astrophysics and the UCLA cluster \textit{Hoffman2} for computational resources. S.N thanks Howard and Astrid Preston for their generous support. B.B. also thanks the the Alfred P. Sloan Foundation and the Packard Foundation for support. M.V. acknowledges support through NASA ATP grants 16-ATP16-0167, 19-ATP19-0019, 19-ATP19-0020, 19-ATP19-0167, and NSF grants AST-1814053, AST-1814259,  AST-1909831 and AST-2007355. NY acknowledges financial support from JST AIP Acceleration Research JP20317829. The research activities described in this paper have been co-funded by the European Union – NextGeneration EU within PRIN 2022 project n.20229YBSAN - Globular clusters in cosmological simulations and in lensed fields: from their birth to the present epoch.

\appendix

\section{Adjusting for different values of $\sigma_8$}\label{sec:appendixSigma8}

\begin{figure}[t]

\fig{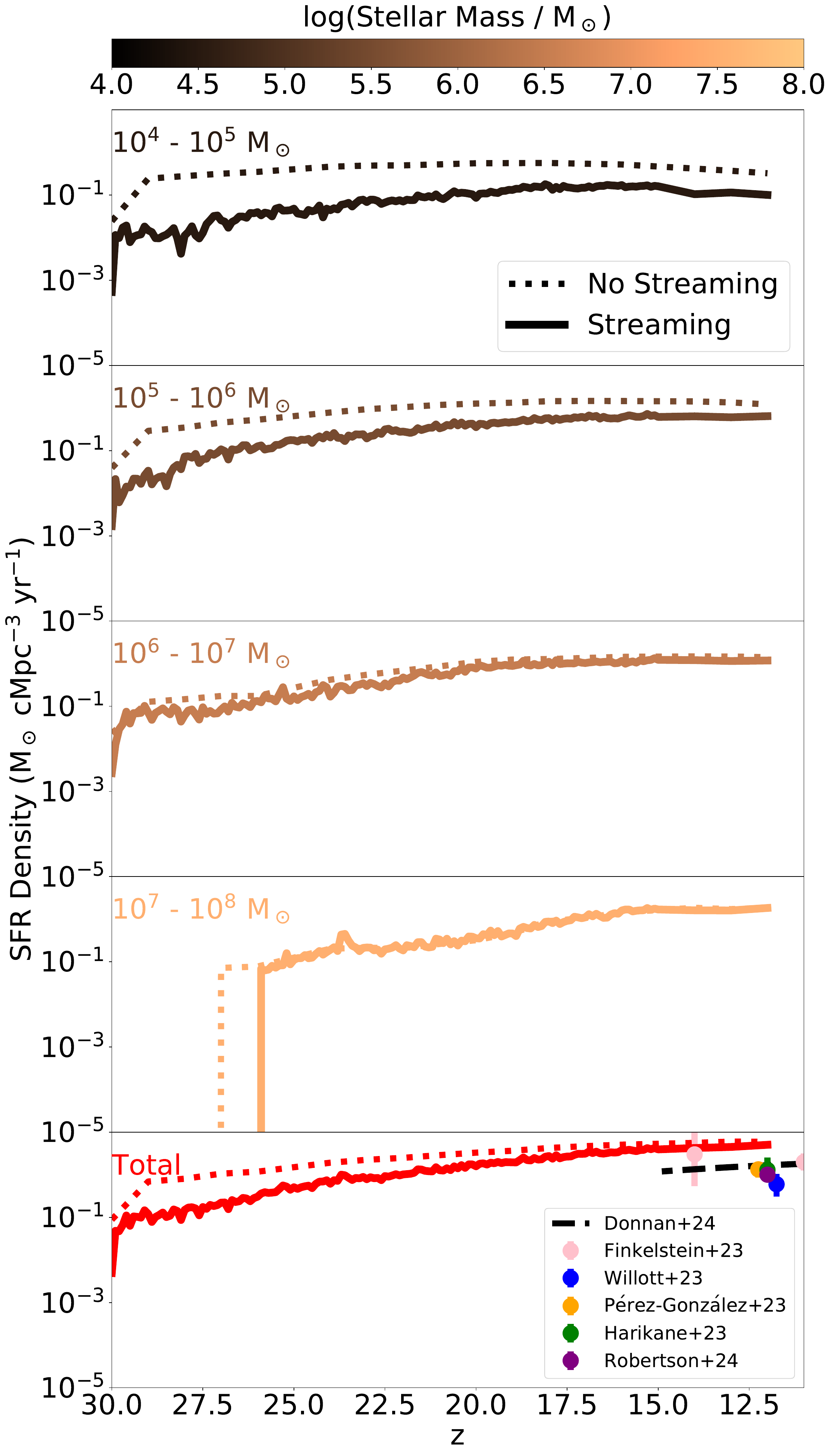}{0.47\textwidth}{}

\caption{{\bf Figure~\ref{Fig:Madau} without adjusting for $\sigma_8$:} This plot shows star formation rate densities with (solid lines) and without (dotted lines) streaming at different object masses, with our enhanced simulation parameter $\sigma_8=1.7$.}\label{fig:MadauUnadjusted}

\end{figure}

As discussed in Section~\ref{ssec:Madau}, we use an enhanced value of $\sigma_8$ in our simulations to increase our statistical power and simulate a rare, overdense region of the Universe. For the purposes of Figure~\ref{Fig:Madau} and Figure~\ref{Fig:sigma8}, we adjust our results to a Universe with different values of $\sigma_8$, using the following analytical correction.

The comoving number density of halos of mass M at redshift z is given by 
\begin{equation} 
\frac{dn}{dM}  = \frac{\rho_0}{M}f_{ST}\left|\frac{dS}{dM}\right| \ , \end{equation} where we use the \citet{Sheth+01} mass function that both fits simulations and includes non-spherical effects on the collapse. The function $f_{ST}$ is the fraction of mass in halos of mass M:
\begin{eqnarray}
 f_{ST}(\delta_c, S) &=&  A' \frac{\nu}{S}\sqrt{\frac{a'}{2\pi}}\left[1+\frac{1}{(a'\nu^2)^{q'}}\right]\exp\left(\frac{-a'\nu^2}{2}\right) \  \\
{\rm and} \quad \nu&=&\frac{\delta_c}{\sigma}=\frac{\delta_c}{\sqrt{S}} \ . \nonumber\end{eqnarray} 
We use best-fit parameters $A'=0.75$ and $q'=0.3$ \citep{Sheth+02}. We can then calculate the relative fraction $f_{0.826}$ of DM halos with said mass M with $\sigma_8=0.826$ compared to $\sigma_8=1.7$ as \begin{equation}\label{Eq:SigmaCorrection}
f_{0.826} = {\left(\frac{dn}{dM}\right)_{\sigma_8=0.826}}{\left(\frac{dn}{dM}\right)^{-1}_{\sigma_8 = 1.7}} \ .
\end{equation}

We apply this adjustment to re-weight the abundances of halos of different masses and match stellar objects to their nearest host halos in order to re-weight these objects in turn (thus producing an adjusted SFR density and number density for each object). Corrections to different values of $\sigma_8$ follow the same process.

\section{Determining density cutoffs for the Kennicutt-Schmidt Relation}\label{sec:appendix}

\begin{figure}[t]

\fig{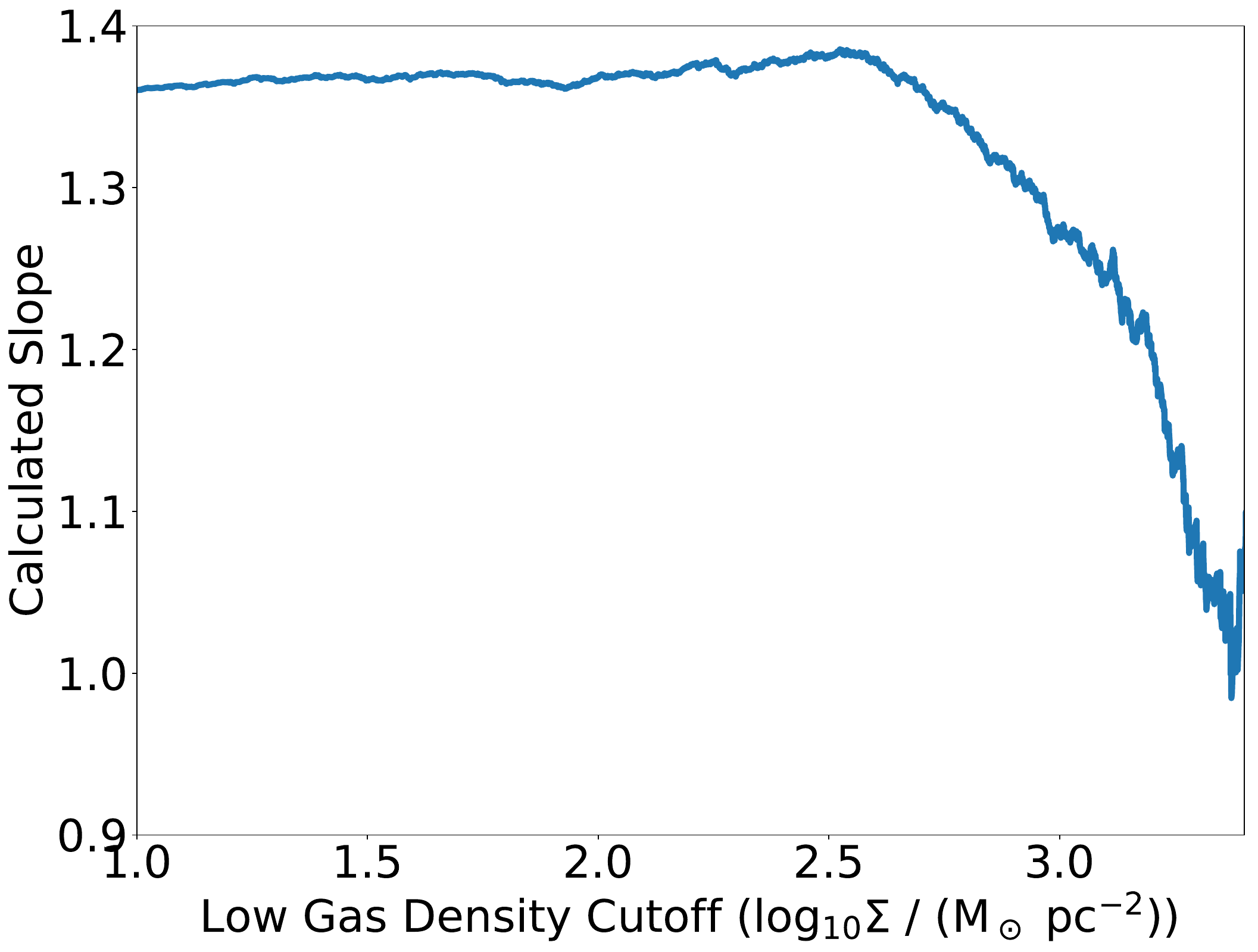}{0.47\textwidth}{\centering Low-density cutoff analysis without streaming to determine a high-density cut}

\caption{{\bf Kennicutt-Schmidt Relation Density Cutoff Selection:} This plot shows the calculated slope of the Kennicutt-Schmidt relation (through a linear regression) as we filter the data with and without streaming using a varying density cutoff (on the x axis). We use the limiting behavior of this plot at high densities to determine an appropriate range of densities over which our results can be taken to be representative of the effect of streaming. Only the more limiting of the two simulations is shown.}\label{fig:KSRDensity}

\end{figure}

In order to calculate the Kennicutt-Schmidt relation, we fit data using the linear relation Equation~\ref{eq:KSR}. Using this and the full data set, we find $\alpha = 1.340\pm0.004$ for the case without streaming and $\alpha = 1.299\pm0.007$ for the case with $2\sigma_{\rm vbc}$ streaming. However, our estimates of the slope of the Kennicutt-Schmidt relation are systematically biased by differing behavior at low, medium, and high gas densities, as well as by the unbalanced populations of low-density star-forming gas objects with and without streaming. This is shown in Figure~\ref{fig:KSRDensity}. As mentioned in Section~\ref{Sec:Halos}, the timescale on which a significant fraction of the gas in the highest-gas-density objects are converted to stars is comparable or less than our timestep between snapshots. This drops the slope of the Kennicutt-Schmidt relation to $1$ at the highest densities, corresponding to complete conversion of gas to stars (up to an efficiency factor). We can calculate the point at which this becomes significant by introducing a low-density cutoff to our data (Figure~\ref{fig:KSRDensity}) and systematically raising it by the gas density intervals present in our data. This allows us to isolate the region of data which is impacted by the changing slope. We take the high-density cutoff to be the point at which the slope began to decrease monotonically (given some unavoidable noise) to $1$, where the errors due to time resolution dominate all other sources of error. This cutoff is calculated to be $\Sigma_{\rm high} = 10^{2.5}$~M$_\odot$~pc$^{-2}$ (rounded to the nearest $0.1$~dex) in the no-streaming case, and $\Sigma_{\rm high, streaming} = 10^{3.0}$~M$_\odot$~pc$^{-2}$ in the less-limiting case with streaming. We select the more limiting of the two cutoffs.

The second complication in our Kennicutt-Schmidt slope is the sparsity of low-density gas objects compared to higher-density gas objects in our data with streaming compared to our data without it. If the Kennicutt-Schmidt relation in our simulation takes on a different slope at low densities, the over-representation of low-density objects in the simulation without streaming would produce a seemingly different slope in a linear KSR fit, even if the behavior of the KSR with compared to without streaming is the same at low and intermediate densities. To minimize this effect, we bin the data into bins of width $0.1$~dex as described in Section~\ref{Sec:Halos}, and calculate a low-density cutoff as the point at which the sparser low-density data set (the simulation with streaming) has $3$ or fewer objects per bin, at which point we have insufficient statistics, which sets our cut at $\Sigma_{\rm Gas} = 10^{0.9}$~M$_\odot$~pc$^{-2}$. Our results then describe and are valid for the intermediate density range $10^{0.9}$~M$_\odot$~pc$^{-2} \leq \Sigma_{\rm Gas} \leq 10^{2.5}$~M$_\odot$~pc$^{-2}$.

\bibliography{myBib}{}
\bibliographystyle{aasjournal}



\end{document}